\documentclass[12pt]{article}
\usepackage{graphics}
\usepackage{bm}
\usepackage{graphicx}
\usepackage{amsmath}
\usepackage{amssymb}
\usepackage{amstext}
\usepackage{amscd}
\usepackage{amsfonts}
\usepackage{appendix}
\usepackage{wasysym}
\usepackage{textcomp}
\usepackage{hyperref}
\usepackage{color}

\DeclareMathAlphabet{\EuFrak}{U}{euf}{m}{n}
\DeclareMathAlphabet{\EuScript}{U}{eus}{m}{n}

\newcommand{\nd}{\noindent}

\newcommand{\be}{\begin{equation}}
\newcommand{\ee}{\end{equation}}
\newcommand{\ben}{\begin{eqnarray}}
\newcommand{\een}{\end{eqnarray}}

\title{{\bf Quantum statistical treatment of Verlinde's conjecture in a Tsallis framework }}
\author{{\small{A. Plastino$^{1,3,4}$, M. C. Rocca$^{1,2,3}$}}, \\
\small{$^1$ Departamento de F\'{\i}sica,
Universidad Nacional de La Plata,}\\
\small{$^2$ Departamento de Matem\'{a}tica,
Universidad Nacional de La Plata,}\\
\small{$^3$ Consejo Nacional de Investigaciones Cient\'{\i}ficas
y Tecnol\'{o}gicas}\\
\small{(IFLP-CCT-CONICET)-C. C. 727, 1900 La Plata -
Argentina}\\\small{$^4$  SThAR - EPFL, Lausanne, Switzerland}}

\date{\today}

\begin{document}

\maketitle

\begin{abstract}
\nd Verlinde has recently conjectured, via a Beckenstein-like thought experiment,  that gravitation, instead of being an elementary force,  is an emergent entropic one.
This rather surprising conjecture was actually    proved  in [Physica A {\bf 505} (2018) 190], in a
strictly classical statistical mechanics' environment. In this Communication, we work in  a quantum statistical context to consider
the conjecture in the case of bosons/fermions, in a Tsallis' framework.   We prove that  Tsallis'
 entropy is the operating potential energy in this quantum treatment, something that does not happen
  in the case  of Boltzmann-Gibbs' entropy. \color{red}  In the classical limit, we show that the emergent force has a Newtonian dependence with the distance. \normalcolor

\vskip 3mm \nd {\bf Keywords} Gravitation, bosons, fermions,
entropic force, emergent force, Verlide's conjecture.\\

\end{abstract}


\newpage

\tableofcontents

\newpage

\renewcommand{\theequation}{\arabic{section}.\arabic{equation}}

\section{Introduction}

\setcounter{equation}{0}

\setcounter{equation}{0}

\nd In 2011, Verlinde \cite{verlinde} conjectured a link between gravity and an entropic force. Such conjecture was
 proved correct recently cite{p1}, in a phase-space, statistical mechanics' context.\vskip 3mm

\nd \color{red} Here we will confront two viewpoints
\begin{itemize}
\item Verlinde's elegant thought experiment, based in black-hole related assumptions, that leads to a Newtonian $r^{-2}$ radial dependence for the entropic force,
 with
\item A quantum statistical mechanics treatment of a Fermi (Bose)  gas that, in its classical limit, leads to the same radial dependence.
\end{itemize}

\nd  \normalcolor Verlinde suggests in his thought experiment  that gravitation should emerge as a result of information about the positions of material particles, connecting a gravity's thermal treatment to 't Hooft's holographic principle. Accordingly, gravitation is to be regarded as  an emergent phenomenon. The idea generated immense attention. See for
instance \cite{times,libro}. A very nice overview regarding the statistical mechanics of gravitation is to be encountered in  Padmanabhan's work \cite{india}, and references therein.\vskip 2mm

\nd This conjecture originated works in cosmology, the dark energy hypothesis, cosmological acceleration, cosmological inflation, and loop quantum gravity. The associated literature is extensive \cite{libro}.  A relevant input is due to Guseo \cite{guseo}, who demonstrated  that the local entropy function, related to a
logistic distribution, is a catenary and vice versa, an invariance interpreted through  Verlinde’s conjecture regarding gravity as an entropic force. \cite{guseo} puts forward a new  interpretation of the local entropy in a system.
 \vskip 2mm

\nd This paper does not deal with any of these issues, though. Considering that we proved Verlinde's conjecture in a classical context \cite{p1}, we wish here to continue a discussion initiated
 in  \cite{DOI} with regards  to the quantal bosons/fermions scenario. In \cite{DOI} we used Boltzmann-Gibbs (BG)  entropy.
 Here we wish to undertake a Tsallis-treatment. Why? Because distinctive  advantages will be accrued in this way.
\color{red}  It turns out that Tsallis entropy is a  potential for the entropic force for $q=4/3$ (see proof in \cite{p1} and also in the Appendix), which is not the BG case.
This makes  Tsallis' entropy the natural information measure to link to gravitation.  This should be natural enough, since it is well known that BG is the natural entropy for systems with short-range interaction, while Tsallis' is the one appropriate to long-range interactions \cite{tsallis}. \normalcolor \vskip 2mm

\nd We base our considerations on Chapters 6 and 7  of \cite{lemons}, to which the reader is referred   for details. \color{red} Only the microcanonical ensemble is used in this book, and thus here. \normalcolor   It is  assumed
that each fermion or boson  possesses an  average energy $E/ N$.
 Such average energy approximation produces results that, while
approximate, describe important features of the ideal Fermi (Bose) gas \cite{lemons}. In fact, most of the book is devoted to this excellent approximation, that allows one to appeal to the micro-canonical ensemble. This entails that the entropy is the logarithm of the multiplicity $\Omega$, according to the celebrated Boltzmann-formula.

\subsection{Our goal}

\nd \color{red} The present effort intends to contribute to the current debate/discussion  regarding
Verlinde's proposal, based on a thought-experiment, for an alternative (entropic) interpretation of gravity. A theory of quantum gravitation does not yet exist. What do we want to achieve here then? We can not expect to obtain en emerging entropic force that will yield classical gravitation in the quantum domain. What we wish to ascertain is whether the classical limit of our quantum statistical mechanics' Verlinde-treatment does yield Newton's gravitation in such limit. We will prove that such is the case. Thus, we contrast Verlinde's thought experiment with a rigorous statistical mechanics' argumentation. Such is the logic of the present effort. \normalcolor

\section{Entropic force for bosons in the microcanonical ensemble}

\subsection{Quantum entropic force}

\setcounter{equation}{0}

We start by reminding the reader of the q-logarithm notion,  defined  according to \cite{tsallis} as
\begin{equation}
\label{eq2.1}
\ln_qx=\frac {x^{1-q}-1} {1-q}.
\end{equation}
\color{blue} Tsallis entropy is defined. for a given set of micro-states labeled by $i$, whose probability is $P_i$,  as \cite{tsallis}

\begin{equation} S_q= -\sum_i\, P_i^q \, \ln_q P_i, \end{equation} with $q$ any real number.
An important portion of the immense Tsallis' literature \cite{tsallis} is devoted to ascertaining which is the appropriate value of $q$ in variegated scenarios. In our present environment we will see below that $q=4/3$. \normalcolor

\vskip 3mm \nd We will use it to compute the multiplicity  $\Omega$ for a Bose gas in the micro-canonical ensemble following Boltzmann's logarithmic prescription in a Tsallis-environment. Following \cite{lemons}, in  a system of free bosons for which the number of  accessible  single-particle states is given by  $n$, $\Omega$  can be thought of as the number of ways of distributing $N$ particles  and $n-1$ ''partitions'' separating them \cite{lemons}. It then  reads  \cite{lemons}
\begin{equation}
\label{eq2.2}
\Omega(E,V,N)=\frac {(N+n-1)!} {N!(n-1)!}=\frac {\Gamma(N+n)} {\Gamma(N+1)\Gamma(n)},
\end{equation}
where the energy $E$, the volume $V$, and the number of bosons $N$ are the extensive variables of the problem at hand. In terms of these variables one has   \cite{lemons}
\begin{equation}
\label{Nn}
\frac {N} {n}=\frac {N} {V}\left(\frac {N} {E}\right)^{\frac {3} {2}}
\left(\frac {3h^2} {4\pi em}\right)^{\frac {3} {2}},
\end{equation}
an important relation that we will often employ below. $m$ stands for the gas' particles' mass. $e$ is Euler's number, and $h$ Planck's constant.
The classical limit is attained for $N/n \ll 1$   \cite{lemons}. Remember also that  the  $\Gamma$ function for large values of  $z$ can be approximated by

\begin{equation}
\label{eq2.4}
\Gamma(z)\approx\sqrt{2\pi}z^{z-\frac {1} {2}} e^{-z},
\end{equation}
which we can use  for
 $N>>1$, $n>>1$ to obtain
\begin{equation}
\label{eq2.5}
\Omega(E,V,N)\approx\frac {(N+n)^{N+n}} {\sqrt{2\pi}N^Nn^n},
\end{equation}
so that the micro-canonical Tsallis' entropy becomes ($k_B$ is Boltzmann's constant)

\begin{equation}
\label{eq2.6}
{\cal S}_q=Nk_B\ln_q\Omega(E,V,N)^{\frac {1} {N}}
\end{equation}
\color{red} For  $q\rightarrow 1$ one has
\begin{equation}
\label{eq1}
{\cal S}=Nk_B\ln\Omega(E,V,N)^{\frac {1} {N}}=
k_B\ln\Omega(E,V,N).
\end{equation}
Thus, for $q\rightarrow 1$  Tsallis's microcanonical  entropy in terms of the multiplicity  becomes  Boltzmann's celebrated one. \normalcolor

\nd Now, \color{blue} it was seen in \cite{p1} that the gravitational interaction can be extracted, 
out of the infinite family of different Tsallis' entropies associated to all possible $q-$values,  only for $q=\frac{4}{3}$.
 \color{red} For other $q-$values the gradient of $S_q$ becomes proportional to $1/r^{\nu}$ with $\nu \ne 2$ \cite{p1}.  See Appendix for details.  \color{blue} Thus, we are heuristically forced to select $q=4/3$. 
Note that, for each value of  $q$, Tsallis has introduced a different statistical mechanics.
For example, for $q=1$ we obtain the orthodox statistical mechanics of Boltzmann-Gibbs. What Tsallis did
it is not to define just  a new (single) realization of statistical mechanics, but a new infinite set of different statistical mechanics' realizations \cite{tsallis}. 
\normalcolor
 Accordingly,
\begin{equation}
\label{eq2.7}
{\cal S}_{\frac {4} {3}}=3Nk_B(1-\Omega^{-\frac {1} {3N}}),
\end{equation}
and using again  (\ref{Nn}) we write
\begin{equation}
\label{eq2.8}
n=V\left(\frac {E} {N}\right)^{\frac {3} {2}}
\left(\frac {4\pi em} {3h^2}\right)^{\frac {3} {2}}.
\end{equation}
It is important to realize that here we find that

\be n \propto V.\label{realize}\ee
It is seen possible at this point  to cast  $\Omega$ in the fashion
\begin{equation}
\label{eq2.9}
\Omega=\frac {e^\gamma} {\sqrt{2\pi}N^N},
\end{equation}
with
\[\gamma=\left[N+V\left(\frac {E} {N}\right)^{\frac {3} {2}}
\left(\frac {4\pi em} {3h^2}\right)^{\frac {3} {2}}\right]
\ln\left[N+V\left(\frac {E} {N}\right)^{\frac {3} {2}}
\left(\frac {4\pi em} {3h^2}\right)^{\frac {3} {2}}\right]-\]
\begin{equation}
\label{eq2.10}
\left[V\left(\frac {E} {N}\right)^{\frac {3} {2}}
\left(\frac {4\pi em} {3h^2}\right)^{\frac {3} {2}}\right]
\ln\left[V\left(\frac {E} {N}\right)^{\frac {3} {2}}
\left(\frac {4\pi em} {3h^2}\right)^{\frac {3} {2}}\right],
\end{equation}
or

\be \gamma= (N+n) \ln{(N+n)} - n\ln{n}. \label{gama}\ee

\vskip 3mm \nd The entropy's gradient becomes then
\begin{equation}
\label{eq2.11}
\vec{\nabla}{\cal S}_{\frac {4} {3}}=k_B\Omega^{-\frac {1} {3N}}
\vec{\nabla}\gamma
\end{equation}
Now, since $V=\frac {4} {3}\pi r^3$ this entails
\begin{equation}
\label{eq2.12}
\frac{\partial{\cal S}_{\frac {4} {3}}} {\partial r}=k_B\Omega^{-\frac {1} {3N}}
\frac {\partial\gamma} {\partial r},
\end{equation}
since ${\cal S}_{\frac {4} {3}}$ depends just upon  $r$. Thus,

\begin{equation}
\label{eq2.13}
\frac{\partial{\cal S}_{\frac {4} {3}}} {\partial r}=k_B
4\pi r^2\Omega^{-\frac {1} {3N}}
\frac {\partial\gamma} {\partial V}.
\end{equation}
Taking into account that
\[\frac {\partial\gamma} {\partial V}=\left(\frac {E} {N}\right)^{\frac {3} {2}}
\left(\frac {4\pi em} {3h^2}\right)^{\frac {3} {2}}\left\{
\ln\left[N+V\left(\frac {E} {N}\right)^{\frac {3} {2}}
\left(\frac {4\pi em} {3h^2}\right)^{\frac {3} {2}}\right]-\right.\]
\begin{equation}
\label{eq2.14}
\left.\ln\left[V\left(\frac {E} {N}\right)^{\frac {3} {2}}
\left(\frac {4\pi em} {3h^2}\right)^{\frac {3} {2}}\right]\right\},
\end{equation}
and remembering Verlinde's definition for the entropic force \cite{verlinde} we find
\begin{equation}
\label{eq2.15}
{\vec {F}}_e=-\lambda k_BT\vec{\nabla}{\cal S}_{\frac {4} {3}},
\end{equation}
or the nice result
\begin{equation}
\label{eq2.16}
F_e=-\lambda k_BT\frac{\partial{\cal S}_{\frac {4} {3}}} {\partial r},
\end{equation}
that by appeal to  (\ref{eq2.14}) yields
\[F_e=-\frac {\lambda} {\beta} 4\pi r^2\Omega^{-\frac {1} {3N}}
\left(\frac {E} {N}\right)^{\frac {3} {2}}
\left(\frac {4\pi em} {3h^2}\right)^{\frac {3} {2}}\left\{
\ln\left[N+V\left(\frac {E} {N}\right)^{\frac {3} {2}}
\left(\frac {4\pi em} {3h^2}\right)^{\frac {3} {2}}\right]-\right.\]
\begin{equation}
\label{eq2.17}
\left.\ln\left[V\left(\frac {E} {N}\right)^{\frac {3} {2}}
\left(\frac {4\pi em} {3h^2}\right)^{\frac {3} {2}}\right]\right\},
\end{equation}
or
\[F_e=-\frac {\lambda} {\beta} 4\pi r^2\Omega^{-\frac {1} {3N}}
\left(\frac {E} {N}\right)^{\frac {3} {2}}
\left(\frac {4\pi em} {3h^2}\right)^{\frac {3} {2}}\left\{
\ln\left[N+\frac {4\pi r^3} {3}\left(\frac {E} {N}\right)^{\frac {3} {2}}
\left(\frac {4\pi em} {3h^2}\right)^{\frac {3} {2}}\right]-\right.\]
\begin{equation}
\label{eq2.18}
\left.\ln\left[\frac {4\pi r^3} {3}\left(\frac {E} {N}\right)^{\frac {3} {2}}
\left(\frac {4\pi em} {3h^2}\right)^{\frac {3} {2}}\right]\right\},
\end{equation}
that can also be cast as
\begin{equation}
\label{eq2.19}
F_e=-\frac {\lambda} {\beta} 4\pi r^2\Omega^{-\frac {1} {3N}}
\left(\frac {E} {N}\right)^{\frac {3} {2}}
\left(\frac {4\pi em} {3h^2}\right)^{\frac {3} {2}}
\ln\left[1+\frac {N} {V}\left(\frac {N} {E}\right)^{\frac {3} {2}}
\left(\frac {3h^2} {4\pi em}\right)^{\frac {3} {2}}\right].
\end{equation}
Note that $F_e$ does not diverge at the origin but vanishes there. However, this happens at distances to the origin of the order of one hundredth of the Planck-length. No practical consequences can be detected, though.
Minding (\ref{Nn}) we also have

\begin{equation}
\label{newFe}
F_e=-\frac {\lambda} {\beta} 4\pi r^2\Omega^{-\frac {1} {3N}}
\left(\frac {E} {N}\right)^{\frac {3} {2}}
\left(\frac {4\pi em} {3h^2}\right)^{\frac {3} {2}}
\ln\left[1+ N/n
\right].
\end{equation}
\color{red} Notice also that the entropic force vanishes at zero temperature and diverges when $T \rightarrow \infty$. This putatively happened at the Big-Bang. There we have $r=0$ as well, so that the behavior of $F_e$ is complicated. However, this does not matter because at these limits quantum gravity, unknown today, reigns.\vskip 3mm

\nd \color{red} A word of caution is necessary here. Since a theory of quantum gravity does not exist yet, we must not naively think that these equations for $F_e$ can be taken at face value. What is really  of interest here is just the classical limit of $F_e$, that we are going to discuss below.     \normalcolor


\subsection{Bose's entropic force in the classical limit  $N/n \ll 1$}

\nd The idea is to judiciously employ (\ref{eq2.19}) and (\ref{newFe}) in this limit. From (\ref{Nn}), i.e.,
  $\frac {N} {n}=\frac {N} {V}\left(\frac {N} {E}\right)^{\frac {3} {2}}
\left(\frac {3h^2} {4\pi em}\right)^{\frac {3} {2}},$ plus $V=4\pi r^3/3$, one sees that

\be \label{2lim}  N/n \ll 1 \,\, \rightarrow \,\, r >> 1.     \ee
From (\ref{Nn}) we also ascertain that we can replace the logarithm by its argument minus unity in (\ref{newFe}), that then becomes

\be F_e=-\frac {\lambda} {\beta} 4\pi r^2\Omega^{-\frac {1} {3N}}
\left(\frac {E} {N}\right)^{\frac {3} {2}}
\left(\frac {4\pi em} {3h^2}\right)^{\frac {3} {2}} \left[\frac {N} {V}\left(\frac {N} {E}\right)^{\frac {3} {2}}
\left(\frac {3h^2} {4\pi em}\right)^{\frac {3} {2}}\right]  \ee
and
\begin{equation}
\label{eq2.20}
F_e=-\frac {\lambda} {\beta} 4\pi r^2\Omega^{-\frac {1} {3N}}
\frac {N} {V}.
\end{equation}
Now, according to (\ref{realize}) we have  $ n \propto V$
and (\ref{eq2.5}) entails that in our limit we have
\be \Omega \propto V^N.   \ee. Thus,
\be \Omega^{-\frac {1} {3N}} \propto (1/V^{1/3}),   \ee.
so that  $F_e$ becomes
\begin{equation}
\label{eq2.22}
F_e \propto -   {r^2},
\end{equation}
 where the proportionality constant is assumed to include Newton's gravitation constant $G$. This
 proves (in statistical mechanics' fashion), for free bosons,  Verlinde's second conjecture: in the classical limit, the corresponding  entropic force decreases as $1/r^2$, like Newton's gravitation  \cite{p1}. \color{red} This dependence of $F_e$ with $r$ is all what Verlinde actually proved in \cite{verlinde}, in a Beckenstein-like thought-experiment. There he {\bf assumes} the number of bits $N$ contained in an appropriate Beckenstein enfolding screen can be cast as $N=Ac^3/G\hbar$, with $A$ the screen's area.
Actually, this is, a priori, Verlinde's definition of $G$, that later will turn out to be gravitation's constant. Summing up, neither in Verlinde's derivation nor in ours we see $G$ emerging from first principles. It is introduced in an  ad hoc fashion, ''by hand''.
    \normalcolor
\section{Entropic force for fermions}

\subsection{Quantum entropic force for fermions}

\setcounter{equation}{0}

Here we deal with $N$ fermions and $n$ micro-states that can be occupied by just one fermion. We have a multiplicity $\Omega$ given by  \cite{lemons}
\begin{equation}
\label{eq3.1}
\Omega=\frac {n!} {(n-N)!N!}=
\frac {\Gamma(n+1)} {\Gamma(n-N+1)\Gamma(N+1)}.
\end{equation}
For  $N>>1$ and $n>>1$ one is allowed to write

\begin{equation}
\label{eq3.2}
\Omega=\frac {e} {\sqrt{2\pi}}\frac {(n+1)^{n+1}}
{(n-N+1)^{n-N+1}(N+1)^{N+1}},
\end{equation}
that can be recast as

\begin{equation}
\label{eq3.3}
\Omega=\frac {e} {\sqrt{2\pi}} e^\gamma,
\end{equation}
where  $\gamma$ is

\begin{equation}
\label{eq3.4}
\gamma=(n+1)\ln(n+1)+(N-n-1)\ln(n+1-N)-(N+1)\ln(N+1).
\end{equation}
The derivative of  $\gamma$ with respect to $V$ is

\[\frac {\partial\gamma} {\partial V}=\left(\frac {E} {N}\right)^{\frac {3} {2}}
\left(\frac {4\pi em} {3h^2}\right)^{\frac {3} {2}}\left\{
\ln\left[1+V\left(\frac {E} {N}\right)^{\frac {3} {2}}
\left(\frac {4\pi em} {3h^2}\right)^{\frac {3} {2}}\right]-\right.\]
\begin{equation}
\label{eq3.5}
\left.\ln\left[1+V\left(\frac {E} {N}\right)^{\frac {3} {2}}
\left(\frac {4\pi em} {3h^2}\right)^{\frac {3} {2}}-N\right]\right\}.
\end{equation}
Retracing  now here the boson-steps  of the preceding Section we find

\[F_e=-\frac {\lambda} {\beta} 4\pi r^2\Omega^{-\frac {1} {3N}}
\left(\frac {E} {N}\right)^{\frac {3} {2}}
\left(\frac {4\pi em} {3h^2}\right)^{\frac {3} {2}}\left\{
\ln\left[1+V\left(\frac {E} {N}\right)^{\frac {3} {2}}
\left(\frac {4\pi em} {3h^2}\right)^{\frac {3} {2}}\right]-\right.\]
\begin{equation}
\label{eq3.6}
\left.\ln\left[1+V\left(\frac {E} {N}\right)^{\frac {3} {2}}
\left(\frac {4\pi em} {3h^2}\right)^{\frac {3} {2}}-N\right]\right\}.
\end{equation}
\color{red} Comparing (\ref{eq2.19}) for bosons with (\ref{eq3.6}) for fermions we see that they are not identical This does not matter, though, since a theory of quantum gravity is not available yet and we should not take the above cited equations at face value. What matters are their classical limits and they do coincide, of course.\normalcolor Finally, we can also cast  (\ref{eq3.6}) as

\[F_e=-\frac {\lambda} {\beta} 4\pi r^2\Omega^{-\frac {1} {3N}}
\left(\frac {E} {N}\right)^{\frac {3} {2}}
\left(\frac {4\pi em} {3h^2}\right)^{\frac {3} {2}}\left\{
\ln\left[1+\frac {4\pi r^3} {3}\left(\frac {E} {N}\right)^{\frac {3} {2}}
\left(\frac {4\pi em} {3h^2}\right)^{\frac {3} {2}}\right]-\right.\]
\begin{equation}
\label{eq3.7}
\left.\ln\left[1+\frac {4\pi r^3} {3}\left(\frac {E} {N}\right)^{\frac {3} {2}}
\left(\frac {4\pi em} {3h^2}\right)^{\frac {3} {2}}-N\right]\right\}.
\end{equation}
Notice that one has, according to the last equation,

\begin{equation}
\label{eq3.8}
1+\frac {4\pi r^3} {3}\left(\frac {E} {N}\right)^{\frac {3} {2}}
\left(\frac {4\pi em} {3h^2}\right)^{\frac {3} {2}}-N>0,
\end{equation}
so that there is a lower bound for $r$ since Eq. (\ref{eq3.8}) entails
\begin{equation}
\label{eq3.9}
r>\left[\frac {3(N-1)} {4\pi}\right]^{\frac {1} {3}}
\left(\frac {N} {E}\right)^{\frac {1} {2}}
\left(\frac {3h^2} {4\pi em}\right)^{\frac {1} {2}}.
\end{equation}

\nd \color{red} Selecting $m=$ uranium's mass, $N$=500, $v=0.1 c$ ($c$=speed of  light) we obtain:
$r=2.4 10^{-21} m$.
This might perhaps suggest a kind of  space-quantization?. Notice also that the entropic force vanishes at zero temperature.
if we select $\lambda$ independent of $T$.\normalcolor

\subsection{Entropic force in the classical limit $N/n \ll 1$}
First of all we realize that (\ref{2lim}) holds in this situation too.
We approximate things now for the classical   limit, starting with (\ref{eq3.7}),  in the fashion
\[F_e=-\frac {\lambda} {\beta} 4\pi r^2\Omega^{-\frac {1} {3N}}
\left(\frac {E} {N}\right)^{\frac {3} {2}}
\left(\frac {4\pi em} {3h^2}\right)^{\frac {3} {2}}\left\{
\ln\left[V\left(\frac {E} {N}\right)^{\frac {3} {2}}
\left(\frac {4\pi em} {3h^2}\right)^{\frac {3} {2}}\right]-\right.\]
\begin{equation}
\label{eq3.10}
\left.\ln\left[V\left(\frac {E} {N}\right)^{\frac {3} {2}}
\left(\frac {4\pi em} {3h^2}\right)^{\frac {3} {2}}-N\right]\right\},
\end{equation}
or

\be
\label{eq3.11}
F_e=\frac {\lambda} {\beta} 4\pi r^2\Omega^{-\frac {1} {3N}}
\left(\frac {E} {N}\right)^{\frac {3} {2}}
\left(\frac {4\pi em} {3h^2}\right)^{\frac {3} {2}} \ln\left[1-\frac {N} {V}\left(\frac {N} {E}\right)^{\frac {3} {2}}
\left(\frac {3h^2} {4\pi em}\right)^{\frac {3} {2}}\right],
 \ee
that also reads

\be F_e= \frac {\lambda} {\beta} 4\pi r^2\Omega^{-\frac {1} {3N}}
\left(\frac {E} {N}\right)^{\frac {3} {2}}
\left(\frac {4\pi em} {3h^2}\right)^{\frac {3} {2}} \ln{[1- N/n]}.
\label{agrego}\ee Expanding now the logarithm we arrive at

\begin{equation}
\label{eq3.12}
F_e=\frac {\lambda} {\beta} 4\pi r^2\Omega^{-\frac {1} {3N}}
\left(\frac {E} {N}\right)^{\frac {3} {2}}
\left(\frac {4\pi em} {3h^2}\right)^{\frac {3} {2}}
\left[-\frac {N} {V}\left(\frac {N} {E}\right)^{\frac {3} {2}}
\left(\frac {3h^2} {4\pi em}\right)^{\frac {3} {2}}\right],
\end{equation}
or, for the entropic force in the classical limit (CL)
\begin{equation}
\label{eq3.13}
F_e(CL) =-(N/V)\frac {\lambda} {\beta} 4\pi r^2\Omega^{-\frac {1} {3N}}.
\end{equation}
\color{red} It is of the essence now to ascertain the behavior of $\Omega$ in the classical limit. Thus, we focus attention upon $\gamma$.
For this, we go back to (\ref{gama}) at this point and realize, that in the classical  limit, it reduces to \normalcolor

\be  \gamma \approx  N \ln{(n)}.\ee
Thus, according to   (\ref{eq3.3}),

\be \Omega \propto \exp{\gamma}= n^N.\ee
Further, we have

\be n \propto V\ee, so that \be \Omega \propto V^N.\ee  Accordingly

\be \Omega^{-\frac {1} {3N}}   \propto V^{-\frac {1} {3}},\label{fomega}\ee
which finally yields, for the entropic force in the classical limit

\begin{equation}
\label{eq3.14}
F_e  \propto -\frac{1}{r^2}.
\end{equation}
 We have encountered a similar expression for the emergent entropic force in the classical
limit similar to that obtained for bosons.

\section{Conclusions}

\nd  Verlinde conjectured in 2011 that gravitation, instead of being an elementary force,  is an emergent entropic one. This rather surprising conjecture had  3280 downloads and 717 cites in ArXiv! In a phase-space classical context it was actually proved true in \cite{p1}.
\vskip 3mm

 \nd Here we asked for its workings in a quantum scenario and it was proved again, this time in the classical limit of the quantum treatment. The relevant question is now: what kind of entropy yields gravitation as an entropic force?
\vskip 3mm

\nd We responded in this work that such an entropy is Tsallis' one for $q=4/3$ \color{blue} (an heuristically choice of $q$),
 \normalcolor  both for fermions and for bosons.

\vskip 3mm

\nd Remark that, for fermions, we have found a lower bound for the distance to the origin $r$.

\newpage

\renewcommand{\thesection}{\Alph{section}}

\renewcommand{\theequation}{\Alph{section}.\arabic{equation}}

\setcounter{section}{0}

\section{Appendix}

\setcounter{equation}{0}

\subsection*{Tsallis'  q-entropy of the free particle}

 Tsallis' q-partition function for a free particle of mass $m$  in $\nu$ dimensions  reads \cite{tsallis}
\begin{equation}
\label{eq5.1}
{\cal Z}_\nu=V_\nu\int \left[1+(1-q)\beta\frac {p^2}
{2m}\right]_+^{\frac {1} {q-1}} d^\nu p,
\end{equation}  with the particle probability distribution $\xi(p)$ being
\be  \xi = \frac{1}{{\cal Z}_\nu}  \left[1+(1-q)\beta\frac {p^2}
{2m}\right]_+^{\frac {1} {q-1}},\ee  where
 $V_\nu$ is the volume of an hypersphere  in $\nu$ dimensions and we assume  $q>1$.
 (\ref{eq5.1}) can be recast as
\begin{equation}
\label{eq5.2}
{\cal Z}_\nu=\frac {2\pi^{\frac {\nu} {2}}}
{\Gamma\left(\frac {\nu} {2}\right)}
V_\nu\int\limits_0^\infty\left[1+(1-q)\beta\frac {p^2}
{2m}\right]_+^{\frac {1} {q-1}} p^{\nu-1}dp.
\end{equation}
With the change of variables  $x^2=\frac {p^2} {2m}$ one has
\begin{equation}
\label{eq5.3}
{\cal Z}_\nu=\frac {(2m\pi)^{\frac {\nu} {2}}}
{\Gamma\left(\frac {\nu} {2}\right)}
V_\nu\int\limits_0^{\frac {1} {(q-1)\beta}}
\left[1+(1-q)\beta x \right]^{\frac {1} {q-1}}
x^{\frac {\nu} {2}-1}dx,
\end{equation}
that after integration becomes
\begin{equation}
\label{eq5.4}
{\cal Z}_\nu=V_\nu\left[\frac {(2m\pi)} {(q-1)\beta}\right]^{\frac {\nu} {2}}
\frac {\Gamma\left(\frac {q} {q-1}\right)}
{\Gamma\left(\frac {q} {q-1}+\frac {\nu} {2}\right)}.
\end{equation}
The mean energy is
\begin{equation}
\label{eq5.5}
<{\cal U}_\nu>=\frac {V_\nu} {{\cal Z}_\nu}  \int \left[1+(1-q)\beta\frac {p^2}
{2m}\right]_+^{\frac {1} {q-1}} \frac {p^2} {2m}d^\nu p,
\end{equation}
or
\begin{equation}
\label{eq5.6}
<{\cal U}_\nu>=\frac {V_\nu} {{\cal Z}_\nu}\frac {(2m\pi)^{\frac {\nu} {2}}}
{\Gamma\left(\frac {\nu} {2}\right)}
\int\limits_0^{\frac {1} {(q-1)\beta}}
\left[1+(1-q)\beta x \right]^{\frac {1} {q-1}}
x^{\frac {\nu} {2}}dx,
\end{equation}
so that after integration we find
\begin{equation}
\label{eq5.7}
<{\cal U}>_\nu=\frac {\nu} {2(q-1)\beta}
\frac {\Gamma\left(\frac {1} {q-1}+\frac {\nu} {2}+1\right)}
{\Gamma\left(\frac {1} {q-1}+\frac {\nu} {2}+2\right)},
\end{equation}
and finally
\begin{equation}
\label{eq5.8}
<{\cal U}>_\nu=\frac {\nu} {[2q+\nu(q-1)]\beta}.
\end{equation}
For the entropy one has  \cite{tsallis}
\begin{equation}
\label{eq5.9}
{\cal S}_\nu=\ln_q{\cal Z}_\nu+{\cal Z}_\nu^{1-q}\beta<{\cal U}>_\nu.
\end{equation}

\subsection*{The Tsallis entropic force}

We specialize things now to
$\nu=3$ and  $q=\frac {4} {3}$.
Why do we select this special value $q=\frac {4} {3}$? There is a solid reason. This is because
$${\cal S}_\nu=\ln_q{\cal Z}_\nu+{\cal Z}_\nu^{1-q}\beta<{\cal U}>_\nu.$$
Since the entropic force is to  be defined as proportional to the gradient of
${\cal S}$, there is a unique $q$-value  for which
the dependence on $r$ of the entropic force is $\sim r^{-2}$
when $\nu=3$. Thus we obtain, for $q=4/3$,
\begin{equation}
\label{eq6.1}
{\cal Z}=\left(\frac {6m\pi} {\beta}\right)^{\frac {3} {2}}
\frac {8\pi}
{\Gamma\left(\frac {11} {2}\right)}r^3,
\end{equation}
\begin{equation}
\label{eq6.2}
<{\cal U}>=\frac {9} {11\beta}.
\end{equation}
Following Verlinde \cite{verlinde} we define the entropic force as
\begin{equation}
\label{eq6.3}
{\vec {\cal F}}_e=-\frac {\lambda(m,M)} {\beta}{\vec {\nabla}{\cal S}},
\end{equation}
where $\lambda$ is a numerical parameter depending on the masses involved, $m$ and a new one $M$ that we place at the center of the sphere. Thus,

\begin{equation}
\label{eq6.4}
{\vec {\cal F}}_e=-\frac {24} {11}
\left[\frac {\Gamma\left(\frac {11} {1}\right)} {8\pi}\right]^{\frac {1} 3}
\left(\frac {k_BT} {6m\pi}\right)^{\frac {1} {2}}
\frac {\lambda(m,M)} {r^2}{\vec e}_r,
\end{equation}
where ${\vec e}_r$ is the radial unit vector.
We see that $F_e$ acquires an appearance quite similar to that of Newton's gravitation, as conjectured by Verlinde en \cite{verlinde}. Note that entropic force vanishes at zero temperature, in agreement with Thermodynamics' third law.

\end{document}